\documentclass[prl,aps,twocolumn,showpacs]{revtex4}
\usepackage{graphicx}
\usepackage{dcolumn}
\newcommand{\nn}{\nonumber}

\begin{document}

\title{Topological Structure of a Vortex in Fulde-Ferrell-Larkin-Ovchinnikov State}

\author{T. Mizushima}
\affiliation{Department of Physics, Okayama University,
Okayama 700-8530, Japan}
\author{K. Machida}
\affiliation{Department of Physics, Okayama University,
Okayama 700-8530, Japan}
\author{M. Ichioka}
\affiliation{Department of Physics, Okayama University,
Okayama 700-8530, Japan}
\date{\today}

\begin{abstract}
We find theoretically that the vortex core in the Fulde-Ferrell-Larkin-Ovchinnikov (FFLO) state is quite different from the ordinary core by a simple topological reason. The intersection point of a vortex and nodal plane of the FFLO state empties the excess spins. This leads to observable consequences in the spatial structure of the spontaneous magnetization. We analyze this topological structure based on the low lying excitation spectrum by solving microscopic Bogoliubov-de Gennes equation to clarify its physical origin.
\end{abstract}


\maketitle

Much attention has been focused on exotic superconductors of various kinds, such as heavy Fermion materials, high $T_c$ cuprates, or organic conductors. These systems are characterized by strongly correlations and/or low-dimensional electronic structures. Here pair potential changes the sign in the momentum space for the relative coordinates of a Cooper pair, such as $p$-, $d$-, $f$-wave symmetries while $s$-wave pairing does not.

The Fulde-Ferrell-Larkin-Ovchinnikov (FFLO) state \cite{ff,lo} differs from these varieties, since the pair potential changes sign in the real space for the center of mass of a Cooper pair. Thus  the superfluid phase in the FFLO state is described by a spatially modulated order parameter. The FFLO state is widely discussed as for its realization in various research fields, ranging from superconductors in condensed matter \cite{machida}, neutral Fermion superfluids in an atomic cloud \cite{mizushima}, to color superconductivity in high energy physics \cite{casalbuoni}.  Originally the FFLO state is proposed when a superconductor is under an applied field where up and down spin populations become unequal by the Zeeman energy shift \cite{ff,lo,gunther}, followed by several theoretical works \cite{tachiki,shimahara,buzdin,adachi}. The excess unpaired carriers, say up spin electrons, are accommodated in the nodal position of the order parameter in space by forming a midgap bound state near the Fermi level \cite{machida}. 

A key factor to stabilize the FFLO state is the relative weight between the Pauli paramagnetic effect and orbital depairing effect due to screening current under a field, apart from the material aspect such as purity and so on. Usually the latter overwhelms the former, thus the upper critical field $H_{c2}$ is orbitally limited. CeCoIn$_5$ \cite{bianchi,tayama,radovan,watanabe,capan,kakuyanagi} is a prime candidate for the strong Pauli effect because of the first order phase transition at $H_{c2}$ in higher fields. In fact, there are several pieces of strong evidence, pointing to the FFLO state in this material \cite{bianchi,tayama,radovan,watanabe,capan,kakuyanagi}.


In order to better characterize the FFLO state and to help its identification, we undertake a detailed study of a vortex state whose topological structure turns out to be quite distinct from conventional flux line. This study is indispensable because the FFLO state may only be induced in vortex state in a superconductor. It is also pointed out that the associated quasi-particle structure around a vortex core is directly probed experimentally via scanning tunneling spectroscopy (STM-STS) through local density of states, and neutron diffraction or X-ray diffraction through the induced magnetization modulation. 

We consider here a single vortex in a superconductor along the $z$-axis with cylindrical symmetry and the order parameter changes its sign at the $z=0$ plane. We solve the Bogoliubov-de Gennes (BdG) equation with the Pauli paramagnetic effect through the Zeeman effect. The Zeeman effect plays a critical role to stabilize the FFLO modulation \cite{ff,lo}. We fully take into account both orbital and spin effects in an equal footing \cite{mizushima,hayashi}. This enables us to reveal an interesting topological structure in a FFLO vortex. 

We start with the BdG equation for the quasi-particle wave functions $u_{\bf q}({\bf r})$ and $v_{\bf q}({\bf r})$ labeled by the quantum number ${\bf q}$:
\begin{eqnarray}
\left[ 
	\begin{array}{cc}
		\mathcal{K} - \mu _{\uparrow} & \Delta ({\bf r}) \\
		\Delta ^{\ast} ({\bf r}) & -\mathcal{K}+ \mu _{\downarrow}
	\end{array}
\right] 
\left[ 
	\begin{array}{c} u_{\bf q}({\bf r}) \\ v_{\bf q}({\bf r}) \end{array}
\right] =  \varepsilon _{\bf q}
\left[ 
	\begin{array}{c} u_{\bf q}({\bf r}) \\ v_{\bf q}({\bf r}) \end{array}
\right],
\label{eq:bdg}
\end{eqnarray}
with $\mathcal{K}=-\frac{\hbar^2\nabla^2}{2m}$ and the self-consistent equation
\begin{eqnarray}
\Delta({\bf r}) = g \sum_{| \varepsilon _q | \le \omega _D} u_{\bf q}({\bf r}) v^{\ast}_{\bf q}({\bf r}) f(\varepsilon _{\bf q}),
\label{eq:gap}
\end{eqnarray} 
where $f(\varepsilon _{\bf q}) = 1/(e^{\varepsilon _{\bf q}/k_BT} + 1)$ is the Fermi-distribution function, and $g$ ($<0$) is the coupling constant. We set the energy cutoff $\omega _{D} = 5\Delta _0$. To prepare the unequal spin population for $\sigma = \uparrow, \downarrow$, the chemical potential is shifted as $\mu _{\uparrow,\downarrow} = \mu \pm \delta\mu$ where the plus (minus) sign corresponds to the up- (down-) spin. We assume the axially symmetric system with a radius $R$ ($=2\xi _0$) and a height $2Z$ ($=5\xi _0$), and isotropic $s$-wave symmetric pairing. Hence, we write the eigenfunctions as $u_{\bf q} ({\bf r}) = u_{\bf q}(r,z) \exp{[i(q_{\theta} - \frac{1}{2})\theta]}$ and $v_{\bf q} ({\bf r}) = v_{\bf q}(r,z) \exp{[i(q_{\theta} + \frac{1}{2})\theta]}$ with $\Delta ({\bf r}) = \Delta (r,z) e^{-i\theta}$ for a single vortex in cylindrical coordinates. $q_{\theta}$ is an angular momentum of the eigenstates. Along the radial and axial directions, the rigid boundary conditions are imposed: $u_{\bf q}(r=R,\theta, z)=v_{\bf q}(r=R,\theta,z)=0$ and $u_{\bf q}(r,\theta, z=\pm Z)=v_{\bf q}(r,\theta,z=\pm Z)=0$. Throughout this paper, $\Delta _0/\varepsilon _F =0.12$, $\delta \mu / \Delta _0 = 0.28$, and $T=0$ are fixed, where $\Delta _0$ is the energy gap in the BCS state at $T\!=\! 0$. We performed full self-consistent computations under these conditions.

\begin{figure}[t!]
\includegraphics[width=0.9\linewidth]{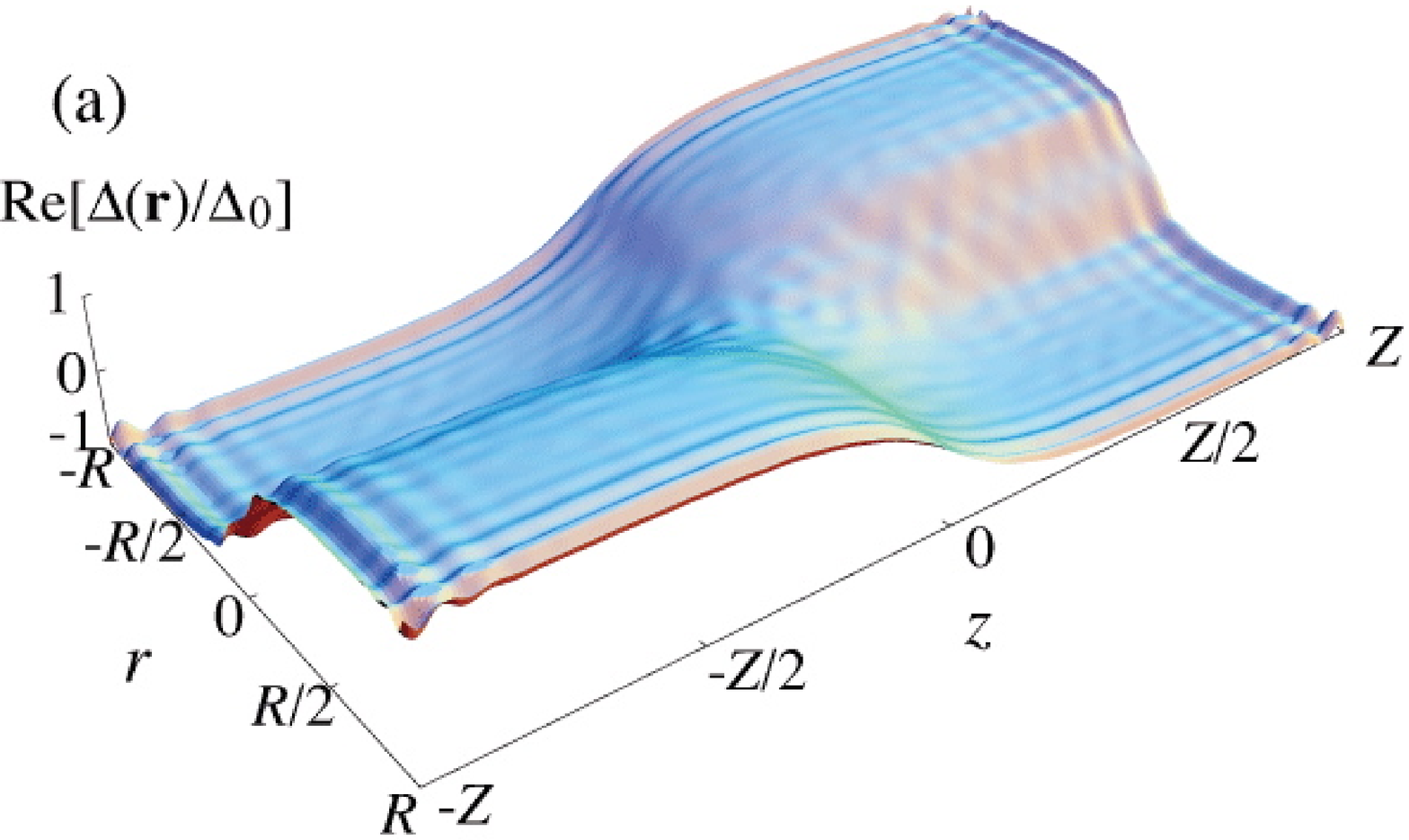} \\
\includegraphics[width=0.9\linewidth]{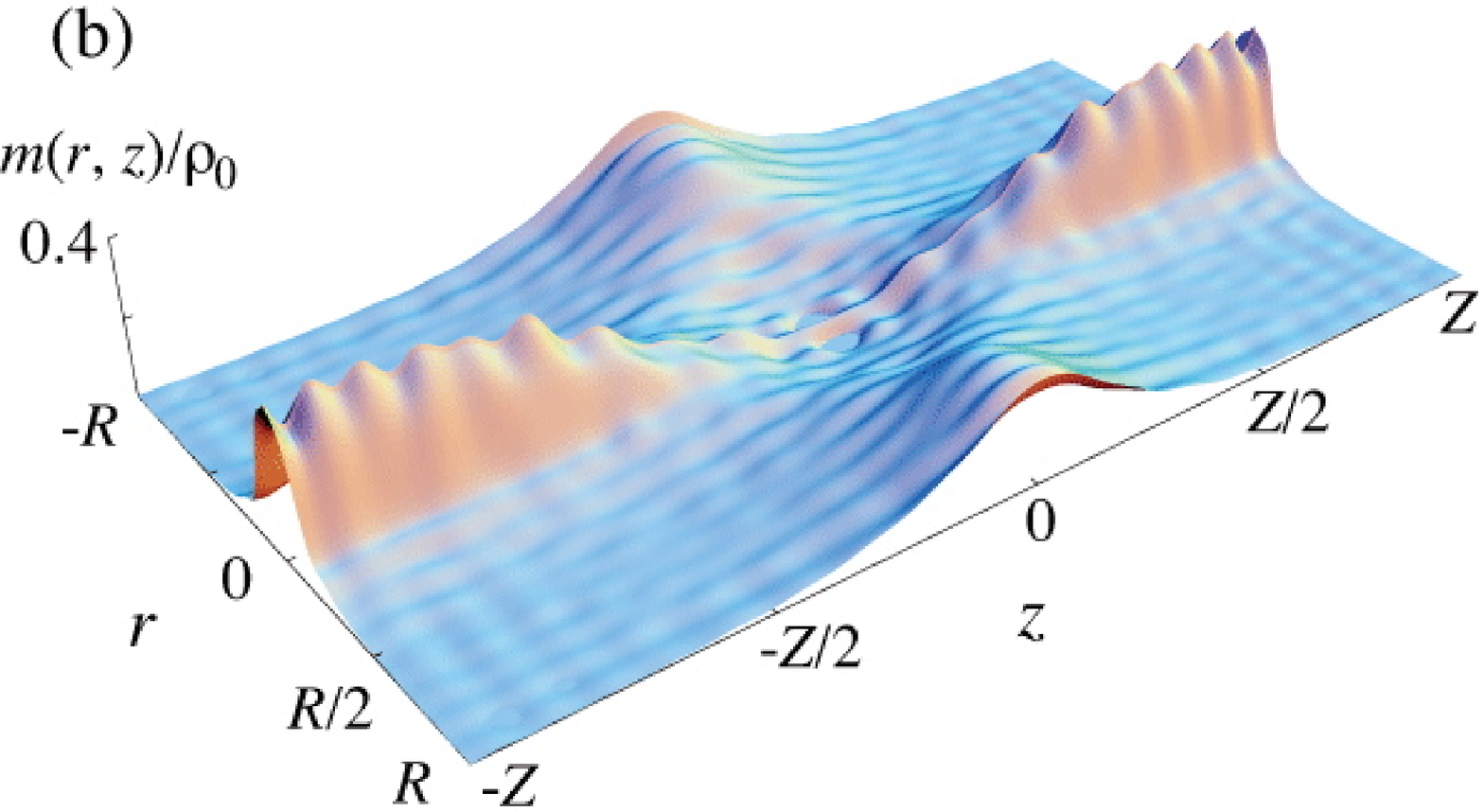}
\caption{
Spatial structures of ${\rm Re}[\Delta({\bf r})/\Delta _0]$ (a) and $m(r,z)/\rho _0$ (b). $\rho _0$ is the density in the uniform BCS state at $T=0$. 
Note that $m(r,z)$ is missing near the origin $r=z=0$.
}
\label{}
\end{figure}

The spatial dependence of the order parameter ${\rm Re}[\Delta({\bf r})]$ is shown in Fig.~1(a), where the vortex line threads at $r=0$ and the nodal plane of the FFLO state is at $z=0$, as shown schematically in Fig.~2. $\Delta({\bf r})$ vanishes on the vortex line at $r=0$ and on the nodal plane at $z=0$ where the sign of $\Delta(r,z)$ changes. It is seen that $\Delta({\bf r})$ approximately varies sinusoidally as a  function of $z$ in this example.

In Fig.~1(b) we show the spatial structure of the spin polarization
$m({\bf r})\!=\! \rho _{\uparrow} ({\bf r}) -\rho _{\downarrow}({\bf r})$, which is defined with the density of each spin component $\rho _{\sigma}$ as 
\begin{eqnarray}
m({\bf r}) = \sum _{\bf q}\left[ |u_{\bf q}({\bf r})|^2f(\varepsilon _{\bf q}) 
- |v_{\bf q}({\bf r})|^2(1+f(\varepsilon _{\bf q})) \right].
\end{eqnarray}
It is seen that $m(r,z)$ is induced both along the vortex core at $r=0$ and on the nodal plane at $z=0$ because the order parameter vanishes there. However, the spin polarization is absent near the origin $(r=0,z=0)$ where the FFLO nodal plane $(z=0)$ and the vortex line $(r=0)$ intersect. This can be understood in terms of the phase structure of the order parameter $\Delta({\bf r})$ around $r=0$ and $z=0$. 

\begin{figure}[b!]
\includegraphics[width=0.85\linewidth]{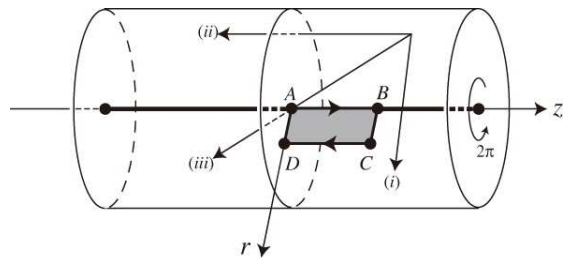} 
\caption{
Schematic view of the vortex line and nodal plane.
}
\label{}
\end{figure}

We display schematically the relative configuration of the vortex line and nodal plane in Fig.~2. Quasi-particles tracing either the path $(i)$ or $(ii)$ feel the sign change of $\Delta({\bf r})$ by cutting through the vortex line with $2\pi$-phase winding $(i)$ or by crossing the nodal plane $(ii)$ (see also Fig.~1(a)), giving rise a bound state near the Fermi level because the order parameter's phase is shifted by $\pi$ on these paths. This is the so-called $\pi$-shift physics \cite{machida}. The former (latter) is the core-localized (node-localized) state. These states accommodate excess up-spin electrons and yield spontaneously the spin polarization at these particular locations. In contrast, the quasi-particles tracing the path $(iii)$, which goes through the origin, do not experience the sign change because of $\pi + \pi$ phase shift, one coming from the vortex contribution and the other from the nodal contribution. Thus it yields no low-energy state due to $\pi$-shift, meaning the absence of the spin polarization at the origin. Thus the spatial structure of the spin polarization shown in Fig.~1(b) is deeply rooted to the topology of the FFLO vortex. Note in fact that these characteristics are absent for the vortex with the winding number 2 ($4\pi$-vortex) by obvious reason ($2\pi+\pi$ shift).

\begin{figure}[b!]
\includegraphics[width=0.9\linewidth]{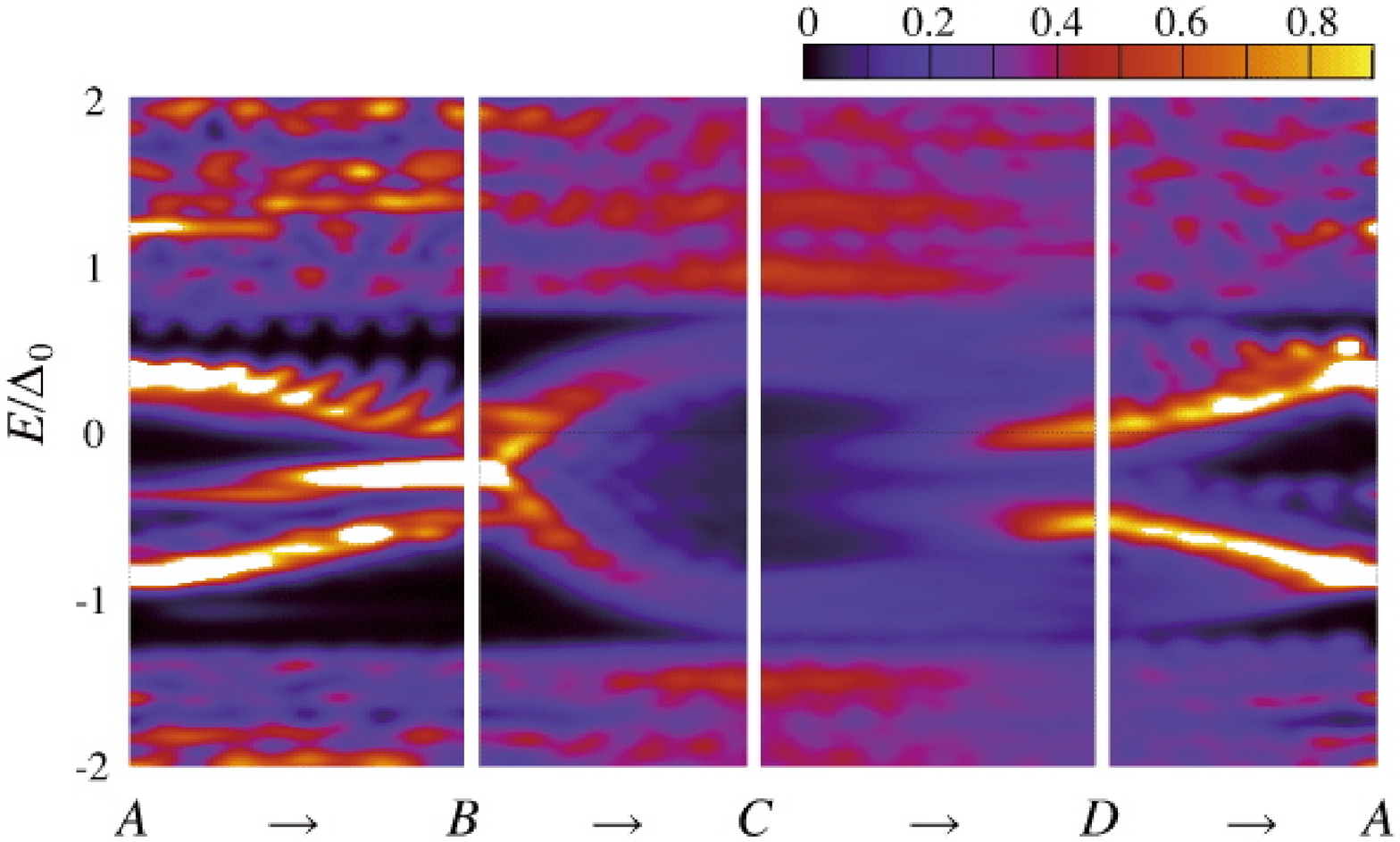} \\
\includegraphics[width=0.9\linewidth]{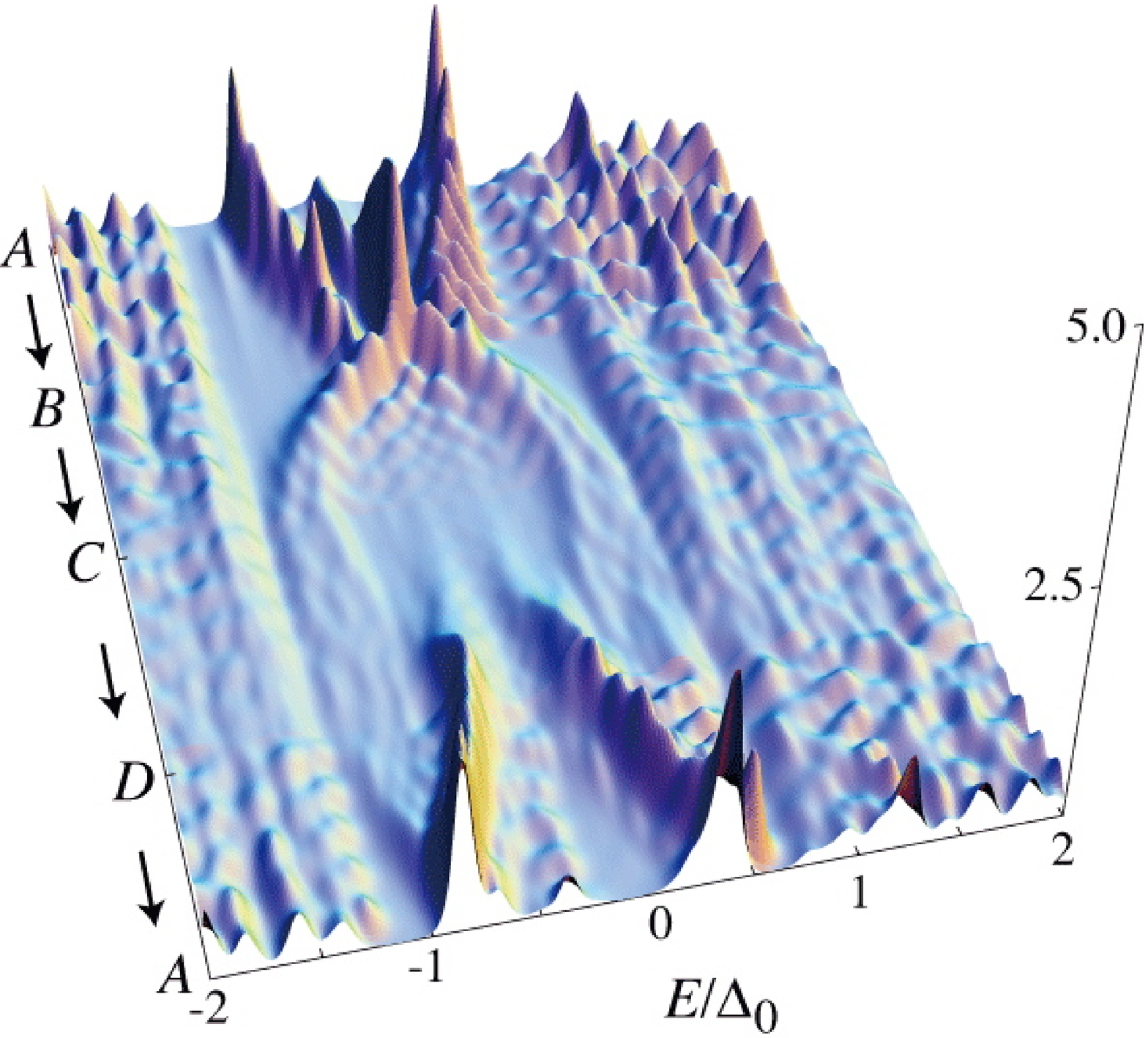} 
\caption{
Spectral evolutions of $N_{\uparrow}({\bf r},E)$ along the paths ($A\rightarrow B\rightarrow C\rightarrow D\rightarrow A$) shown in Fig.~2. Density plot (upper) and stereographic view (lower) are displayed. In the energy scale, $E\!=\! 0$ corresponds to the Fermi energy $\mu$.
}
\label{}
\end{figure}

We now consider the electronic structure associated with the FFLO vortex, namely the local density of states (LDOS), $N_{\sigma}({\bf r},E)$ which is calculated by 
\begin{eqnarray}
&& N_{\uparrow}({\bf r},E) = \sum _{\bf q} |u_{\bf q}({\bf r})|^2 \delta(E - \varepsilon _{\bf q}),  \nn \\
&& N_{\downarrow}({\bf r},E) = \sum _{\bf q} |v_{\bf q}({\bf r})|^2 \delta(E+\varepsilon _{\bf q}).
\end{eqnarray}
In Fig.~3, we show the spectral evolution $N_{\uparrow}({\bf r},E)$ along the paths $A(r=0,z=0)\rightarrow B(0,Z/2)\rightarrow C(R/2,Z/2)\rightarrow D(R/2,0)\rightarrow  A$ presented in Fig.~2. We plot the spectrum $N_{\uparrow}({\bf r},E)$ and $N_{\downarrow}({\bf r},E)$ at the position $A$, $B$, and $D$ in Fig.~4. The energy gap opens in $|E-\mu _{\uparrow}| \alt \Delta _0$ in $N_{\uparrow}({\bf r},E)$. Outside the gap, the continuum spectra are seen. At the vortex core outside the FFLO node (Fig.~4(a)), $N_{\uparrow}({\bf r},E)$ and $N_{\downarrow}({\bf r},E)$ have, respectively, the zero-energy peak at $\mu _{\uparrow}$ and $\mu _{\downarrow}$ due to the core-localized state. In the path $B \rightarrow C$ in Fig.~3, we see a typical spectral evolution found in the conventional vortex \cite{hayashi}. As we move away from the core site $B$ towards $C$, the higher angular momentum state with the higher energy acquires the spectral weight. On the other hand, in the nodal plane of the FFLO state, the peak of the core states for up (down) spins at $\mu _{\uparrow}$ ($\mu _{\downarrow}$) is split into two. Along the vortex line $A \rightarrow B$ in Fig.~3, we see that the split core-localized bound states tend to merge, forming a usual core bound state at $B$. Away from the core site $A$ towards $D$ along the nodal plane, the splitting of the two peaks decreases. In the FFLO node without vortices, these two peaks are merged at $E = \mu _{\uparrow}$. 

\begin{figure}[t!]
\includegraphics[width=0.85\linewidth]{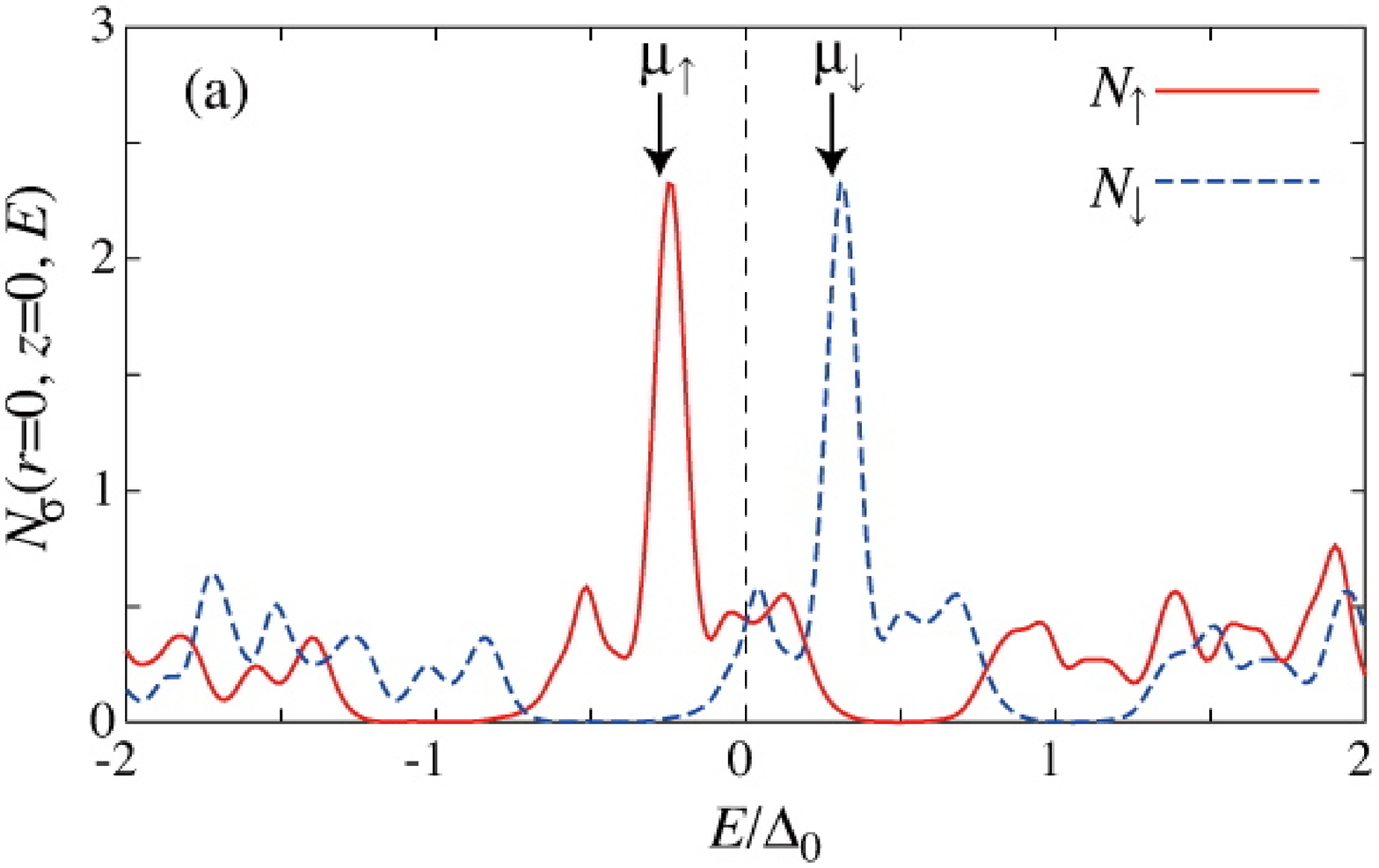} \\
\includegraphics[width=0.85\linewidth]{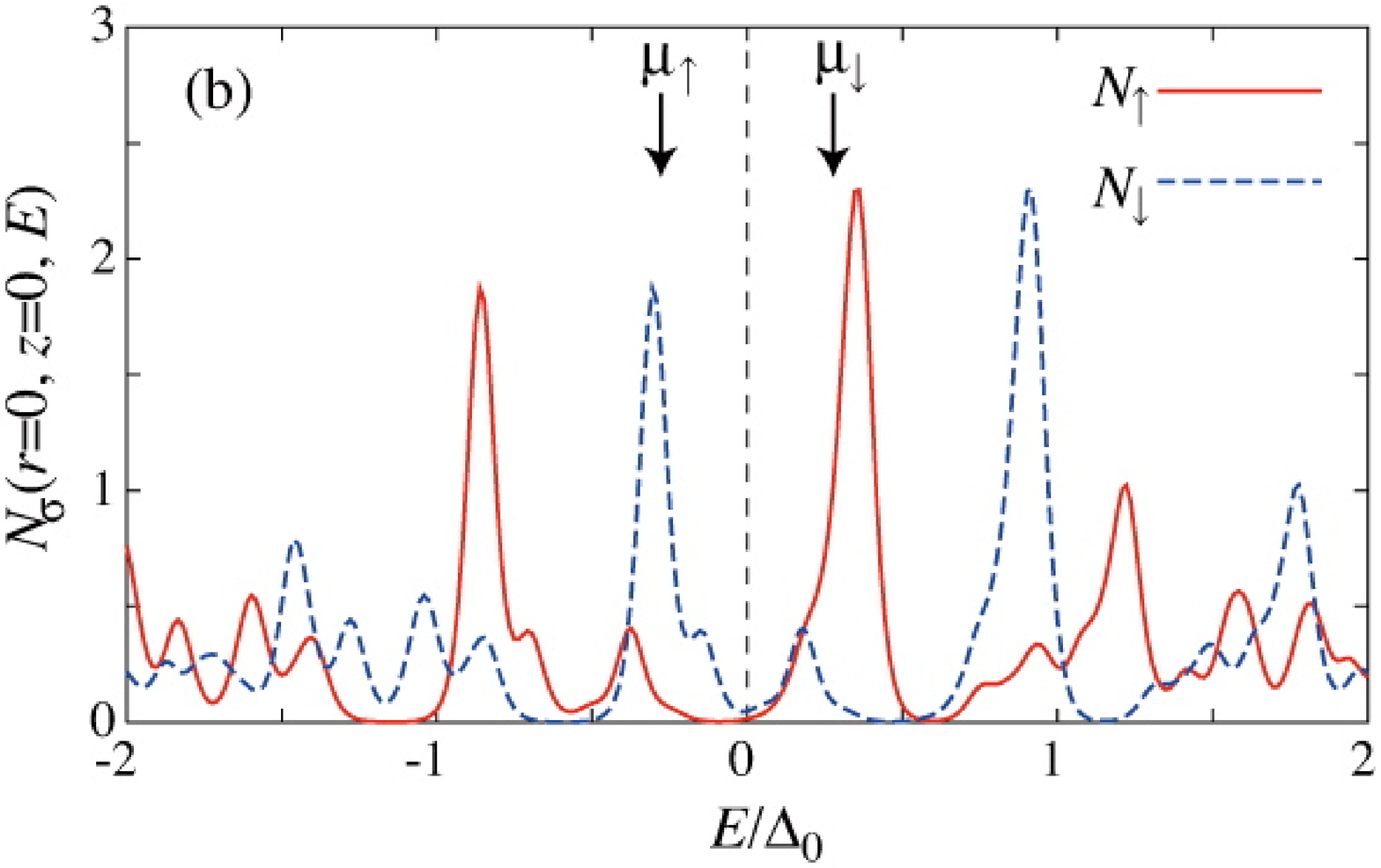} \\
\includegraphics[width=0.85\linewidth]{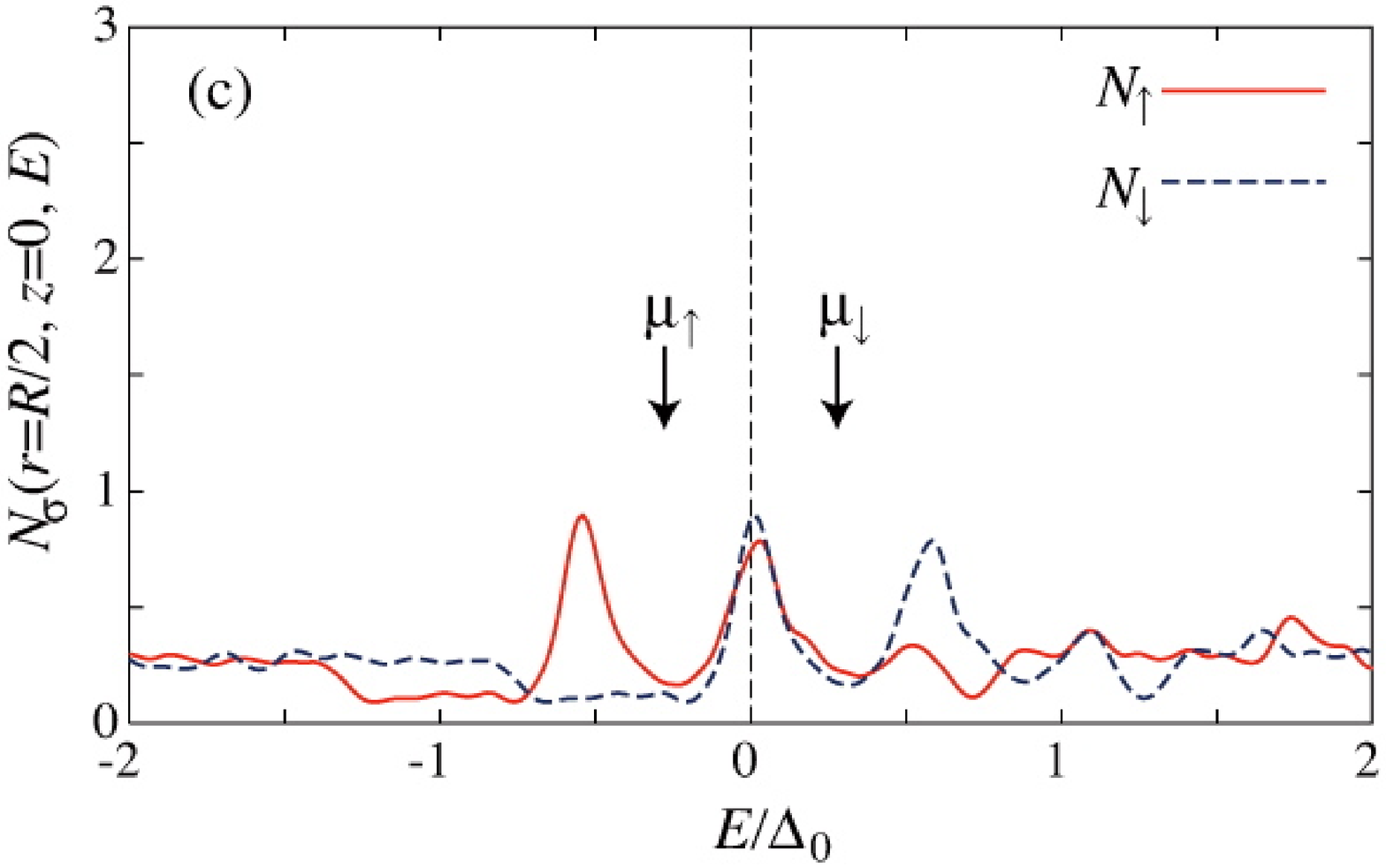} 
\caption{
The local density of states for spin up (solid lines) and spin down (dashed lines) components at the positions $B$ (a), $A$ (b), and $D$ (c).
}
\label{}
\end{figure}

The LDOS structures $N_{\uparrow}({\bf r},E)$ and $N_{\downarrow}({\bf r},E)$ are closely related to the local spin polarization $m({\bf r})$ in Fig.~1(b). 
At the vortex core (Fig.~4(a)) or at the nodal plane (Fig.~4(c)), most of the peak states for up (down) spins are located at $E<0$ ($E>0$), and they are occupied (empty). Thus, the spin polarization appears. However, at the vortex core in the nodal plane (Fig.~4(b)), since the one of the split peaks is occupied both for up and down spins, the up and down spin populations are equal. This is the reason from the electronic structure why the spin population is absent in the vortex core on the nodal plane. 

Obviously STM-STS is a direct method to see these spectral evolutions in the FFLO state. In particular, comparison of tunneling conductance at $A$ and $D$ shown in Fig.~4 yields direct evidence for the FFLO state.

The origin of the splitting of the vortex core bound state at the position $A$ is due to the Doppler shift by the FFLO modulation. If we assume $\Delta({\bf r})=\Delta _0 e^{\pm i Q_z z}$ to simply catch the essential feature, the wave functions can be written as $u_{\bf q}({\bf r})=u_{\bf q}(r,z)e^{\pm i Q_z z/2}$ and $v_{\bf q}({\bf r})=v_{\bf q}(r,z)e^{\pm i Q_z z/2}$ which satisfy Eqs.~(1) and (2). Thus, in the kinetic term $\mathcal{K}$ in Eq.~(1), $\nabla^2_z \rightarrow \nabla^2_z \pm iQ_z \nabla _z - Q^2_z/4$. Since the quasi-particles traveling across the nodal plane with the finite $z$-component $v_{F,z}$ of the Fermi velocity are important to form the bound states at the nodal plane, the Doppler shift term $\pm i Q_z \nabla _z \propto Q_z v_{F,z}$ gives the upper or lower energy shift of the spectrum, producing the split of the peaks for the vortex core states. We also calculate the LDOS spectrum using the test potential with $\Delta(r,z)=\Delta(r)\sin{(Q_z z)}$. It is seen that with increasing $Q_z$, the splitting of the peak becomes wider.

We should comment on the NMR experiment \cite{kakuyanagi} of CeCoIn$_5$ where ``novel'' high field phase for $H \parallel c$ is identified as the FFLO state. Our result is basically consistent with the data: Double peaks in the resonance spectrum are related to the spin polarization structure. One peak is associated with the nodal plane of the FFLO state, the other is for the bulk far from the nodal plane. Detailed calculations show that we obtain a double-peak shape in the magnetization distribution profile, corresponding to the Knight shift in the NMR experiment. 

We propose here neutron diffraction experiment on the FFLO state. If we look at a scattering plane $(0,q_y,q_z)$ of the reciprocal space $(H,K,L)$ for $H_{\rm external} \parallel z$, there is no magnetic Bragg diffractions at $(0,0,\pm Q_z)$ because the scattering vectors are parallel to the spin polarization ($\parallel H$). However, around the higher Bragg points at $(0,2\pi n/L,\pm Q_z)$ ($L$ is the vortex lattice spacing, $n=1,2,3, \cdots$) we have the spin polarization component perpendicular to the scattering vectors, which can be observable in this particular setting for diffraction experiments. Note that the vortex lattice spacing $L \sim 200 {\rm \AA}$ for 1T and $Q^{-1}_z$ is a long period ($Q^{-1}_z$ changes widely depending on $H$ and $T$). That must be small angle scatterings. The scattering images the spin polarization, consisting of the sheets with the period of $Q_z$ along the field direction and rod-like objects along vortex lines with voids of the polarization at the intersection. This spatial modulation of the spin polarization can be detected by synchrotron X-ray diffraction via magneto-striction, where the lattice modulation is caused by the coupling to the spin modulation in the FFLO nodal plane.

As mentioned before, the present results are independent of the detailed material parameters because the topological structure in the FFLO  vortex is essential. As for the pairing symmetry; We assume here isotropic $s$-wave pairing. The $d$-wave pairing which is advocated for CeCoIn$_{5}$ \cite{aoki} does not alter our conclusions based on the $\pi$-shift of the FFLO state. In that case, the extra-nodal excitations associated with vortex core at the Fermi energy extend towards the nodal directions of the anisotropic pairing function \cite{ichioka}.

We remark that the present considerations can be applicable to resonance Fermionic superfluid under rotation with mismatched Fermi surfaces of two species in ultracold atoms where vortices are created and observed \cite{mit}.

In summary, we have examined the vortex core structure in the FFLO state in a superconductor under applied field. By a simple topological reason, the spin polarization of excess up-spin electrons vanishes around the place where the vortex line intersects with the nodal plane, both of which accommodate excess up-spin electrons. This intricated spatial magnetization profile may be probed by neutron or X-ray diffraction experiments as extra-diffraction spots, or NMR and $\mu$SR experiments through the resonance line shape. This topological explanation is confirmed by detailed computations based on the BdG equation. We provide full account of these features through analyses of the LDOS at various places. These spectral features should be directly checked by STM-STS experiments. We have proposed these experiments in connection with the ``novel'' high field phases reported in CeCoIn$_5$.
The present conclusion is also hold for ultra-cold atomic Fermion condensates with unequal two species populations under rotation \cite{mizushima}. 
 
We thank  H. Adachi, Y. Matsuda, E. M. Forgan  and H. Yoshizawa for useful discussions.

\end{document}